\documentclass[journal]{IEEEtran}
\usepackage{cite}

\ifCLASSINFOpdf
  \usepackage[pdftex]{graphicx}
\else
\fi
\usepackage{amsmath}

\usepackage{stfloats}
\usepackage{siunitx}
\usepackage{tabularx}
\usepackage[dvipsnames]{xcolor}

\begin{document}
%
\title{A Silicon Photonic Neural Network for Chromatic Dispersion Compensation in 20 Gbps PAM4 Signal at 125 km and Its Scalability up to 100 Gbps}
%
%
%

\author{Emiliano~Staffoli,
        Gianpietro~Maddinelli,
        and~Lorenzo~Pavesi,~\IEEEmembership{Fellow,~IEEE}
\thanks{E. Staffoli, G. Maddinelli, and L. Pavesi are with Department
of Physics, University of Trento, Trento,
38123, Italy, e-mail: emiliano.staffoli@unitn.it.}
\thanks{Manuscript received XXXXX X, X; revised XXXXX X, X.}
}

%
%

\markboth{Journal of Lightwave Technology,~Vol.~X, No.~X, XXXXX~XXXX}%
{Shell \MakeLowercase{\textit{et al.}}: Bare Demo of IEEEtran.cls for IEEE Journals}
%



\maketitle

\begin{abstract}
A feed-forward photonic neural network (PNN) is tested for chromatic dispersion compensation in Intensity Modulation/Direct Detection optical links. The PNN is based on a sequence of linear and nonlinear transformations. The linear stage is constituted by an 8-tap time-delayed complex perceptron implemented on a Silicon-On-insulator platform and acting as a tunable optical filter. The nonlinear stage is provided by the square modulus of the electrical field applied at the end-of-line photodetector. The training maximizes the separation between the optical levels (i.e. the eye diagram aperture), with consequent reduction of the Bit Error Rate. Effective equalization is experimentally demonstrated for 20 Gbps 4-level Pulse Amplitude Modulated signal up to 125 km. An evolutionary algorithm and a gradient-based approach are tested for the training and then compared in terms of repeatability and convergence time. The optimal weights resulting from the training are interpreted in light of the theoretical transfer function of the optical fiber. Finally, a simulative study proves the scalability of the layout to larger bandwidths, up to 100 Gbps. 
\end{abstract}


%
\IEEEpeerreviewmaketitle

\section{Introduction}
%
%
%
%
\IEEEPARstart{I}{n} the last decades, with the advent of the Internet data transmission assumed a central role in the development of society. Fiber optics-based communications represented a breakthrough technology in this field, allowing for long-distance transmission at large bandwidths. Nowadays, this technology represents the backbone of many applications, including cloud services, e-commerce, streaming platforms, big data transfer inter- and intra-data centers, and, recently, Artificial Intelligence \cite{zhu2020first,Cheng2018recent,wei2020special,kumar2020intra}. Market's needs require an always higher Capacity $\times$ Distance product in the transmission lines, pushing the research to find new transmission and detection techniques that adapt to different scenarios and costs \cite{karar2023short,minkenberg2021copackaged}. Coherent transceivers are mostly employed in long-haul propagation, where maximum spectral efficiency is required \cite{Renaudier2015spectrally}. To increase the propagation distance, high optical launch power can be used. This triggers linear and nonlinear effects \cite{agrawal2010fiber_ch2}, which in turn induce transmission impairments. Error compensation is provided by Digital Signal Processing (DSP) performed at the transceiver level \cite{Faruk2017digital}, often requiring Application-Specific Integrated Circuits (ASICs) featuring high power consumption and introducing latency \cite{kuschnerov2014energy}. On the other hand, short-reach applications rely on Intensity Modulation/Direct Detection (IM-DD), providing lower spectral efficiency but reducing hardware complexity, costs, and energy consumption compared to the coherent counterpart \cite{zhu2020comparative}. In this framework, errors in data transmission are induced by linear effects such as Polarization Mode Dispersion, Symbol Timing Offset, Optical filtering, and Chromatic Dispersion (CD) \cite{agrawal2010fiber_ch2}. The latter represents the dominant source of impairments, causing a time broadening of the propagating pulses with the consequent Intersymbol Interference (ISI). This effect becomes more severe with increasing bandwidth and propagation distance. Dispersion Compensating Fibers \cite{gruner2000dispersion} are nowadays widely used to counteract ISI, with their application being affected by the high production costs, their additional contribution to the latency in the transmission, and their non-tunability (being these static devices after the implementation). The alternative of Chirped Fiber Bragg Gratings \cite{hill1994chirped} drastically reduces the latency, but high power consumption is required for thermal or electrical tuning of the devices. A further option is represented by integrated Photonic Neural Networks (PNNs) \cite{chaoran2022prospects} which allow to overcome many of the limitations just mentioned. Signal processing performed directly in the optical domain reduces power consumption and minimizes latency. Moreover, being tunable, a PNN can be adapted to different transmission scenarios by tuning the $\beta_m$ parameters that define its dispersion profile ($\beta_m = d^m \beta/d\omega^m$, $\beta$ being the single-mode propagation constant \cite{agrawal2010fiber_ch2}). Finally, PNNs realized as Photonic Integrated Circuits (PICs) in CMOS-compatible processes feature low costs and small footprints, allowing the embedding of the photonics platforms directly into a transceiver \cite{shekhar2024roadmapping}.

In this work, we experimentally demonstrate the use of an integrated PNN for CD equalization. The \textcolor{black}{very simple} design is based on an 8-channel time-delayed complex perceptron \cite{mancinelli2022photonic}. The input optical signal is split into 8 copies that accumulate a relative time delay \textcolor{black}{$\Delta t = 25$ ps} between each other. Each copy is then applied with an amplitude and a phase weight and finally recombined with the others. The present design is a feed-forward architecture, with the nonlinear activation function constituted by the square modulus of the propagating electric field applied at the end-of-line photodetector. Signal processing is performed with minimized latency at the optical level, except for the training phase, which is performed offline. Signal equalization is applied to a 10 Gbaud 4-level Pulse Amplitude Modulated (PAM4) signal propagating in a standard single-mode fiber up to 125 km long. \textcolor{black}{ A first demonstration of the use of a time-delayed complex perceptron was reported in \cite{staffoli2023equalization}. There, a PNN with 4 channels, longer delay lines ($\Delta t = 50$ ps), and only phase weights was tested. Here, we report on an improved PNN with more channels and shorter $\Delta t$. These ensure that more pieces of optical information closer in time interact with each other. A further improvement is represented by the use of also amplitude weights, which allow for better handling of the various delayed optical contributions. In the previous design, all the delayed copies of the input reached the recombination stage, leaving the phase weights only to arrange the interference. This possibly introduces unnecessary information from the past to the recombination stage.}

This paper is organized as follows: Section \ref{sec:procedures} introduces the PNN layout, the experimental setup, and the training procedures, focusing in particular on the loss function definition and the choice of the minimization algorithm. The corresponding results are reported and discussed in Section \ref{sec:conclusion}, followed by an exhaustive comparison of the presented technology with other approaches in Section \ref{sec:comparison}. Finally, Section \ref{sec:conclusion} addresses future developments of our design in terms of the number of taps and delay units for its adaptability to higher modulation frequencies up to 100 Gbps.

\section{Experimental Setup and Procedures}
\label{sec:procedures}
\subsection{Delayed complex perceptron layout}
\label{sec:dcp_layout}

\begin{figure}[t]
    \centering
    \includegraphics[width=1\linewidth]{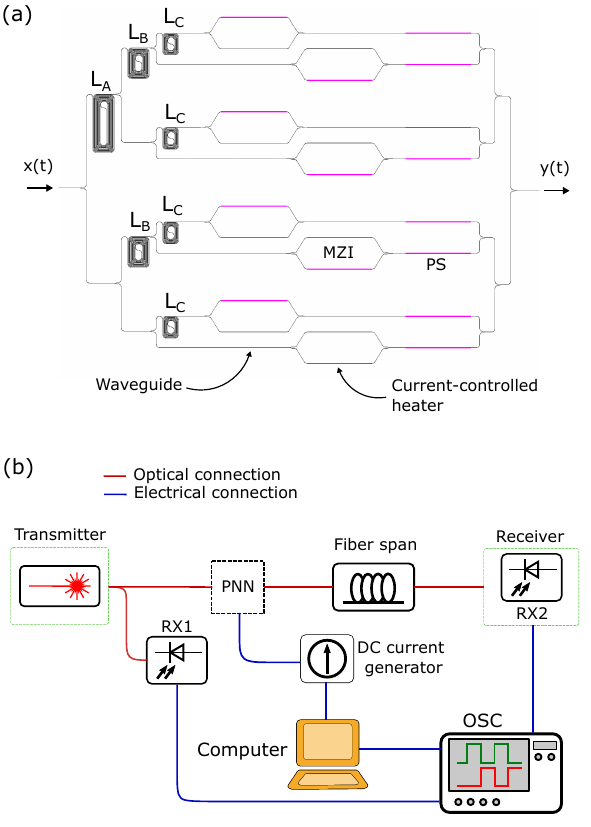}
    \caption{(a) Layout of the 8-channel time-delayed complex perceptron (PNN device). The input signal $x(t)$ propagates through a cascade of y-branches and spiralized optical paths producing 8 time-delayed copies hosted in as many parallel waveguides. The spirals are available in 3 lengths $L_A = 7.10$ mm, $L_B = 3.55$ mm, and $L_C = 1.77$ mm, corresponding to an added propagation time respectively of $\Delta t_A = 100$ ps, $\Delta t_B = 50$ ps, and $\Delta t_C = 25$ ps. The delayed copies are then applied with an amplitude and a phase weight via a series of a Mach-Zehnder interferometer (MZI) and a phase shifter (PS) hosted in each channel. These are controlled via current-driven micro-heaters placed on top of the waveguides. Finally, the output signal is obtained by complex-summing the weighted copies via a $1\times 8$ combiner realized with a cascade of y-branches. (b) Simplified experimental setup. The modulated optical signal is sent to the PNN device used as a pre-compensator for CD effect. The processed signal propagates then through a fiber span with variable lengths, between 0 km and 125 km with steps of 25 km. Two fast photodiodes connected to an oscilloscope (OSC) acquire respectively the input (RX1) and output (RX2) traces, then send them to the computer for loss function evaluation. A DC current generator drives the weights applied by the PNN device.}
    \label{fig:simplified_setup}
\end{figure}

Fig.  \ref{fig:simplified_setup}(a) shows the design of the PNN device under test. It is fabricated in a Silicon-on-Insulator platform, and made by Si waveguides with a $220 \times 500$ nm$^2$ cross-section surrounded by Silica cladding. The input and output ports are located respectively at the leftmost and rightmost edges of the chip. The input signal $x(t)$ enters the chip via butt coupling and propagates through a cascade of 50/50 y-branches and spiralized optical paths. Overall, this creates $N=8$ time-delayed copies of the input signal, ending in as many parallel waveguides (channels). Spirals are available in 3 different lengths, respectively $L_A = 7.10$ mm, $L_B = 3.55$ mm, and $L_C = 1.77$ mm, each corresponding to an added propagation time of $\Delta t_A = 100$ ps, $\Delta t_B = 50$ ps, and $\Delta t_C = 25$ ps. The channels are labeled from the first ($i=1$) at the bottom edge to the eighth ($i=8$) at the top edge. 
The $i$-th channel hosts a delayed copy $x_i(t) = x[t - (i-1)\Delta t]$ of the input signal, where the delay is measured with respect to the first channel. The quantity $\Delta t = \Delta t_C$ is the delay unit imposed by the shortest spirals between copies traveling in adjacent channels. The eighth channel is associated with the maximum delay of $\Delta t_A + \Delta t_B + \Delta t_C = 7 \Delta t$. Each channel hosts a Mach-Zehnder Interferometer (MZI) with a footprint of $ 400 \times 150 \, \mu$m$^2$, and two y-branches provide the initial splitting and final recombination. Each MZI is controlled via the thermo-optic effect produced by a current-driven TiN micro-heater placed on top of one of the arms. The final aperture level (i.e. the transmittance) of the MZI determines the amplitude weight $a_i$ applied to the optical signal traveling in that channel. A phase weight $e^{j\phi_i}$ is then generated for each channel by another current-controlled micro-heater (phasor, PS), leading to a weighted optical signal $x[t - (i-1)\Delta t] k_i a_i e^{j\phi_i}$ where $k_i$ accounts for the channel losses (numerical values provided in Appendix \ref{sec:experimental}). The weighted copies are then complex-summed via a $8\times 1$ combiner realized via a cascade of 50/50 y-branches providing the output sequence
\begin{equation}
    y(t) = \sum_{i=1}^{N=8} x[t - (i-1)\Delta t] k_i a_i e^{j\phi_i}.
    \label{eq:y_cperc}
\end{equation}
Therefore, the PNN device acts as an 8-tap Optical Finite Impulse Response filter, and the currents driving the MZIs and the PSs represent the tunable parameters of the PNN. The theoretical free spectral range (FSR) is $1/\Delta t= \SI{40}{\giga\Hz}$. The number of taps $N_T$ is determined starting from the empirical formula already presented in \cite{staffoli2023equalization}:
\begin{equation}
    N_T = \text{int} \left( \frac{1/B + \vert L \beta_2 \Delta \omega \vert }{\Delta t} \right).
    \label{eq:ntaps}
\end{equation}
Here, the numerator represents an estimate of the increased time width of a Gaussian pulse after propagation in a fiber of length $L$, obtained as the sum of the initial baud time slot $1/B$ (with $B$ being the baud rate) and the pulse broadening $\Delta T=|L\beta_2\Delta\omega|$ induced by CD. We consider the Group Velocity Dispersion parameter $\beta_2 = -0.021$ ps$^2$/m of a standard SM G.652D fiber, and $\Delta\omega$ is the pulse bandwidth. The first generation design tested in \cite{staffoli2023equalization} featured phase weights only, with $B = 10$ Gbps, $L = 100$ km, and $\Delta t = 50$ ps, resulting in  $N_T = 4$. The $\Delta t$ value was chosen to have a sufficiently dense sampling of the information in a single-bit time slot (at least 2 samples per baud). For the current design, $\Delta t = 25$ ps is chosen, which allows dealing with multilevel formats and faster data rates. Indeed, this layout has approximately the same maximum relative delay (given by $[N -1] \times \Delta t$) between the channels as the previous one but with a finer sampling of the information from the same baud simultaneously sent to the recombination stage. Summarizing, the selected $N$ and $\Delta t$ are a trade-off between a sufficiently dense optical sampling, insertion loss minimization, and a reduced number of trainable parameters. Finally, the design of the photonic structures implemented on the PNN device is optimized for the Transverse Electric (TE) mode propagation, making the PNN device polarization sensitive.

\subsection{Ideal design working principle}

The frequency response of an optical fiber with length $L$ on the envelope of the input field is \cite{gliese1996chromatic}
\begin{equation} \label{eq:of_transfun}
    H_{of}(\omega) = e^{j\beta(\omega_0+\omega)L},
\end{equation}
where $\beta(\omega)$ is the frequency dependent propagation constant of the fiber and $\omega_0$ is the optical carrier frequency. By expanding $\beta(\omega)$ around $\omega_0$ up to the second order, \eqref{eq:of_transfun} can be approximated as
\begin{equation} \label{eq:of_transfun_exp}
    H_{of}(\omega) = e^{j\left( \beta_0 + \beta_1\omega + \frac{1}{2}\beta_2\omega^2 \right)L}.
\end{equation}
Since we are interested in CD, $\beta_0$ and $\beta_1$ can be removed from the expression as they affect the signal with a constant phase shift and a propagation delay respectively, therefore
\begin{equation} \label{eq:of_transfun_beta2}
    H_{of}(\omega) \approx e^{j\frac{1}{2}\beta_2L\omega^2}.
\end{equation}
Equation \eqref{eq:of_transfun_beta2} shows that CD affects the propagation by applying a phase mask to the different frequency components of the signal. Perfect CD compensation can be achieved by a device with the transfer function
\begin{equation}
    H(\omega) = e^{-j\frac{1}{2}\beta_2L\omega^2},
\end{equation}
such that 
\begin{equation} \label{eq:Hideal}
    H(\omega)H_{of}(\omega)=1.
\end{equation}
The impulse response of such a device is
\begin{equation}
    h(t) = e^{-j\frac{1}{2\beta_2L}t^2},
\end{equation}
and the output $y(t)$ is the convolution of the input $x(t)$ with the impulse response
\begin{equation} \label{eq:y_inf_cperc}
    y(t) = \int_{-\infty}^{\infty} \mathrm{d}\tau\, x(t-\tau) e^{-j\frac{1}{2\beta_2L}\tau^2}.
\end{equation}
Let us note that according to \eqref{eq:y_cperc} the PNN performs a discretized convolution of its input over a finite window of time. Comparing \eqref{eq:y_cperc} with \eqref{eq:y_inf_cperc}, it emerges that perfect CD compensation can be achieved by a delayed complex perceptron featuring infinite taps ($N\to\infty$), an infinitesimal delay unit ($\Delta t\to0$) and lossless channels ($k_i=1$). Furthermore, all MZIs are to be kept open ($a_i=1$), and the PSs must induce a relative phase equal to
\begin{equation} \label{eq:phases_ideal}
    \phi_i = -\frac{1}{2\beta_2L}(i-N/2)^2\Delta t^2.
\end{equation}
referred to the central channel ($i=N/2$). \textcolor{black}{However, real device implementations have a limited number of taps and a finite delay unit, making them non-ideal. Therefore, the optimal phases provided by a training procedure are supposed to depart from those given by \eqref{eq:phases_ideal}}. 


\subsection{Experimental setup overview and loss function definition}
The PNN device has been tested with the experimental setup presented in a simplified version in Fig.  \ref{fig:simplified_setup}(b). An extensive description of the experimental setup can be found in Appendix \ref{sec:experimental}. In the transmission stage, a laser source operating at 1550 nm is modulated as a 10 Gbaud PAM4 signal based on periodic Pseudo-Random Binary Sequences of order 11, producing an overall period of $2^{10}$ symbols. A Fiber Optic Coupler sends part of the optical signal to a 20 GHz-bandwidth fast photodiode (RX1), where the input signal is acquired for reference. The other fraction proceeds to the PNN for optical processing, as described in \eqref{eq:y_cperc}. The amplitude and phase weights are regulated by an external DC current generator. Then, the output signal from the PNN is sent through an optical fiber span with variable length, from 0 km (Back-To-Back configuration, BTB) to 125 km with steps of 25 km. The position of the PNN device within the transmission line makes it, in fact, a pre-compensator for CD effects. Placing the PNN device before the propagation stage allows for getting rid of polarization diversity effects since the device is polarization-sensitive. \textcolor{black}{ The optical power in the setup remains low enough to avoid nonlinear effects and damage in any component. In particular, it stays below 3 dBm at the input of the PNN and below 2 dBm at the input of each optical fiber span.} After fiber propagation, the optical signal is detected by another 20 GHz bandwidth fast photodiode (RX2). Both RX1 and RX2 are connected to a 16 GHz bandwidth oscilloscope with a sampling frequency of 80 GSa/s. 

For each measure (i.e. transmission test) the DC current generator sends an array of pre-set currents to the PNN device to drive the amplitude and phase weights. A triggering signal is then sent to the oscilloscope to acquire the resulting periodic trace at the end-of-line receiver. The output trace is then aligned via the cross-correlation method with the digital target sequence obtained as a periodic version of the digital input sequence. Each aligned output trace $y$ serves as an input to a properly defined loss function that assesses the quality of the transmission, which is then provided to the minimization algorithms for the PNN training. This is performed first via a Particle Swarm Optimizer (PSO) \cite{kennedy1995particle} and then via the Adam algorithm \cite{kingma2017adam}, namely a gradient-based approach provided with memory to minimize the possibility of a premature end of the research in a local minimum. 

The amount of errors in a single data transfer is quantified by the Bit Error Rate (BER), here defined as the sum of the discrepancies between the binary de-mapped version of the digitized target and output sequences. However, the discrete-valued nature of the BER function makes it not well-suited for a gradient-based approach to minimization. In addition, the minimum BER value of 0 can be obtained statistically even in sub-optimal equalization conditions, possibly leading the PSO algorithm to end the research prematurely. A BER optimization can be equivalently obtained by minimizing the overlap between the distributions of samples associated with each expected optical level, as described by (2) in \cite{staffoli2023equalization}. Thus, we proceed with defining a function that measures the separation between the distributions. 

\begin{figure*}
    \centering
    \includegraphics[width=\linewidth]{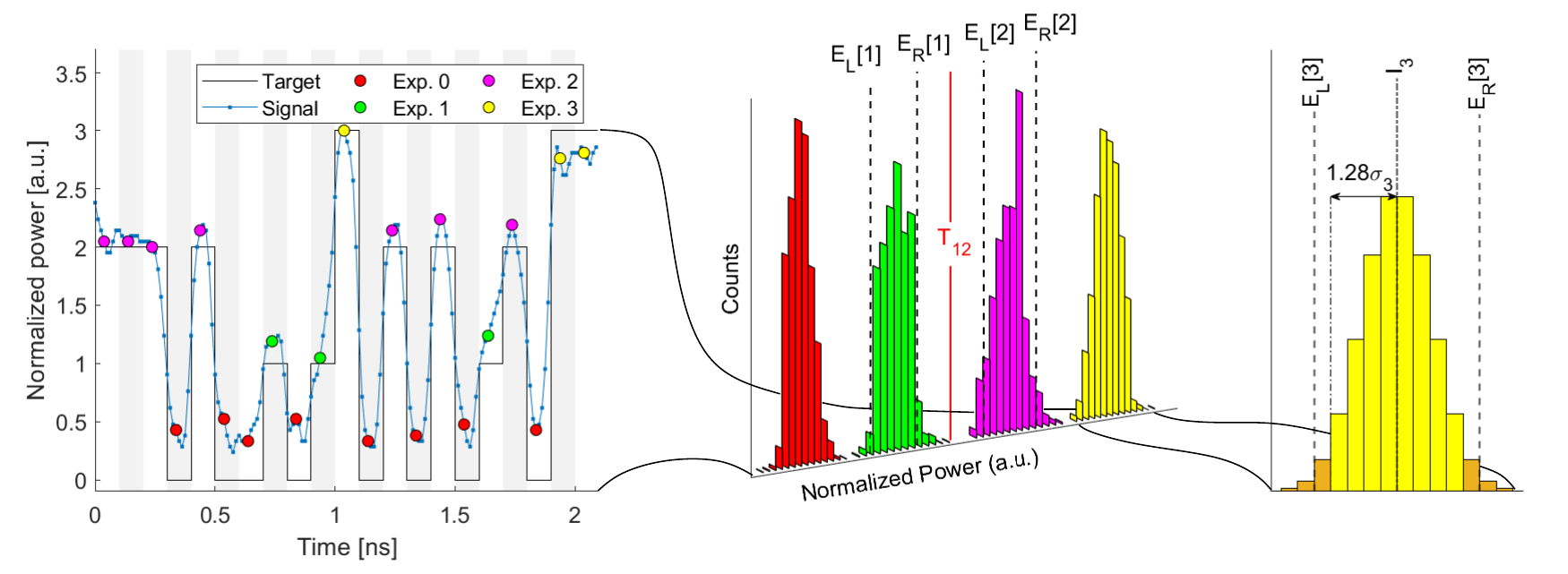}
    \caption{Illustration of the one-sample separation loss function $\mathcal{L}_1(k)$ evaluation. The current representation refers to $k=N_{sps}/2$, and any reference to it in subscripts or superscripts has been omitted for clarity. The left panel presents a time trace of the output signal $y$ (blue line) as acquired by RX2, with 8 samples per symbol (blue dots), and the software-generated target signal (black line). The sub-sampling of $y$ at the $k$-th sample generates $y_k$, whose samples (large colored circles) are grouped into $\{y\}_{k,n}$ depending on the expected level $n$, with $n=0,\dots,3$. Samples in each group are distinguished by color (red, green, purple, yellow) and populate the corresponding distributions presented in the central panel. The left and right tails' position $E_L[n]$ and $E_R[n]$ for each distribution enter in the definition of $\mathcal{L}_1$ (see \eqref{eq:s1}), while the thresholds $T_{n,n+1}$ (see \eqref{eq:threshold}) produce a digitized version of $y_k$. For the sake of clarity, the visual representation is limited to the quantities referred to the two central levels. The right panel analyzes a single distribution ($n=3$ characterized by its mean value $I_n$ and its standard deviation $\sigma_n$. The value of $E_L[n]$ ($E_R[n]$) results from the average position of the leftmost (rightmost) 10\% of the population, both represented by orange bars. }
    \label{fig:hist_example}
\end{figure*}

Fig. \ref{fig:hist_example} provides a visual representation of the procedures leading to the loss function evaluation. The measurement produces a $N_s = 7.1 \times 10^4$ samples-long output trace $y$ (blue line in Fig. \ref{fig:hist_example}) with $N_{sps} = 8$ samples per symbol because the 10 Gbaud signal is sampled at 80 GSa/s. A sequence $y_k$ (colored large dots) is then obtained by sub-sampling $y$ at the $k$-th sample in each symbol, with $k=N_{sps}/2$. The sub-sampling is performed close to the center of the symbol, namely where the contrast is ideally the highest, allowing retrieving the maximum information from each symbol. The points in $y_k$ are grouped into $\{y\}_{k,n}$ based on their expected level $n=0,\dots,N_L-1$ ($N_L$ being the number of levels in the chosen PAM format) determined by comparison with the target sequence. In Fig. \ref{fig:hist_example} the points belonging to each $\{y\}_{k,n}$ are marked with a different color and generate the distributions shown in the central panel. Each distribution is populated by $N_n$ samples $\{y^i\}_{k,n}$, with $i=1,\dots,N_{n}$. The rightmost panel focuses on a single histogram, showing the typical Gaussian envelope identified by its average $I_{k,n}$ and variance $\sigma_{k,n}$. The position of the left and right tails, respectively $E_L[k,n]$ and $E_R[k,n]$, can be estimated as
\begin{equation} \label{eq:E_L_def}
    E_L[k,n] =  \frac{1}{n_L} \bigg \vert \sum_{i=1}^{n_L} \{{y}^i\}_{k,n} \bigg\vert ,
\end{equation}
\begin{equation} \label{eq:E_R_def}
    E_R[k,n] =  \frac{1}{n_R} \bigg \vert \sum_{i=1}^{n_R} \{y^i\}_{k,n} \bigg\vert.
\end{equation}
In \eqref{eq:E_L_def}, the index $i$ is such that $\{y^i\}_{k,n} < I_{k,n} - 1.28\sigma_{k,n}$, namely the sum is performed over the $n_L$ points corresponding to the 10\% of the population in the leftmost area of the distribution. Analogously, in \eqref{eq:E_R_def} $i$ is such that $\{y^i\}_{k,n} > I_{k,n} + 1.28\sigma_{k,n}$, namely the sum is performed over the $n_R$ points corresponding to the 10\% of the population in the rightmost area of the distribution. These areas are identified by the orange bars in the third panel of Fig.  \ref{fig:hist_example}. We define the one-sample separation loss function as
\begin{equation}
    \mathcal{L}_1(k) = \underset{n=0,\dots,N_L-2}{\text{max}} \big\{ E_R[k,n] - E_L[k,n+1] \big\},
    \label{eq:s1}
\end{equation}
which returns negative values when all the distributions are correctly separated. A minimization of $\mathcal{L}_1$ produces the maximum separation between the optical levels, that is the maximum aperture of the eye diagram. Since the aperture level is measured for a single sample (the $k$-th) close to the center of the symbol, the minimization may lead to a local separation of the levels, losing the symmetry in the eye diagram aperture around the center of each symbol. Therefore, it is convenient to define a two-sample separation loss function as
\begin{equation}
    \mathcal{L}_2 (k) = \mathcal{L}_{1}(k) + \mathcal{L}_{1}(k+1).
\end{equation}
This measures the separation between the optical levels in two different points within each symbol, respectively symmetrical to its center. A minimization of $\mathcal{L}_2$ reduces the risk of asymmetric eye diagrams both on the horizontal and vertical axis. The definitions of $\mathcal{L}_1$ and $\mathcal{L}_2$ represent a generalization to a multi-level system of the separation loss function presented in \cite{staffoli2023equalization}. Finally, the BER evaluation starts with the digitization of the output trace. The separation threshold $T_n$ between the $n$-th and the $n+1$-th level are defined as
\begin{equation}
    T_{n,n+1} = \frac{1}{2} \left( E_R[k,n] + E_L[k,n+1] \right).
    \label{eq:threshold}
\end{equation}
with $n = 0,\dots,N_L-2$. These are then applied to $y_k$ producing the corresponding 4-level digitized sequence $y_{dig}$ which is then de-mapped into the binary domain $y_{bin}$. The BER is then evaluated as an error counting between $y_{bin}$ and the corresponding de-mapped binary target sequence $\Bar{y}_{bin}$. Each evaluation is performed on $2 \times N_s/N_{sps} = 1.8\times 10^4$ bits, providing a minimum BER value of $5.5\times10^{-5}$.

\subsection{Dispersion-induced power penalty}
In IM-DD links, CD is known to cause the fading of specific frequencies of the transmitted signal \cite{devaux1993simple}. This effect is briefly reviewed in this section. Let $s_{in}(t)$ be the input signal to an optical fiber of length $L$. According to \eqref{eq:of_transfun_beta2}, the output signal $s_{out}(t)$ can be written as
\begin{equation}
    s_{out}(t) = \int\mathrm{d}\omega\,S_{in}(\omega)e^{j\frac{1}{2}\beta_2L\omega^2} e^{-j\omega t},
\end{equation}
where $S_{in}(\omega)$ is the Fourier transform of $s_{in}(t)$. Assuming that $s_{in}(t)=a+b\cos(\bar\omega t)$,
\begin{equation}
    s_{out}(t) = a + be^{j\frac{1}{2}\beta_2L\bar\omega^2}\cos(\bar\omega t).
\end{equation}
The signal detected at the input is proportional to
\begin{equation}
    |s_{in}(t)|^2 = a^2 + 2ab\cos(\bar\omega t) + b^2\cos^2(\bar\omega t),
\end{equation}
while the signal detected at the output is proportional to
\begin{equation}
    |s_{out}(t)|^2 = a^2 + 2ab\cos(\beta_2L\bar\omega^2/2)\cos(\bar\omega t) + b^2\cos^2(\bar\omega t).
\end{equation}
The CD-induced penalty of the transmission line at $\bar\omega$ is defined as the ratio between the frequency component at $\bar\omega$ of $|s_{in}(t)|^2$ and that of $|s_{out}(t)|^2$ \cite{gliese1996chromatic},
\begin{align}
\begin{split}
    P(\bar\omega) &= 20\log_{10}\left| \frac{\mathcal{F}\{|s_{in}(t)|^2\}(\bar\omega)}{\mathcal{F}\{|s_{out}(t)|^2\}(\bar\omega)} \right| \\
    &= -20\log_{10}\left|\cos(\beta_2L\bar\omega^2/2)\right|,
\end{split}
\label{eq:pp}
\end{align} 
where $\mathcal{F}$ denotes the Fourier transform. The formula predicts that the first notch of the CD-induced penalty is located at $\omega_m=\sqrt{\pi/\beta_2L}$. As long as this value is larger than the bandwidth of the modulation, the effects of CD are minimal. When this condition is not satisfied, CD severely impairs the transmission.

Experimentally, the CD-induced penalty of a transmission line is measured by modulating the intensity of the laser with a cosine of frequency $\bar\omega$. In our experiment the value of $\bar\omega$ is swept from \SI{50}{\mega\Hz} to \SI{15}{\giga\Hz} taking steps of \SI{50}{\mega\Hz}. The result of the measurement is expected to follow \eqref{eq:pp} when the PNN is not inserted in the transmission line. On the other hand, when the PNN compensates CD, the power penalty is expected to be flat within the bandwidth of the signal.

\begin{figure*}[t]
    \centering
    \includegraphics[width=\linewidth]{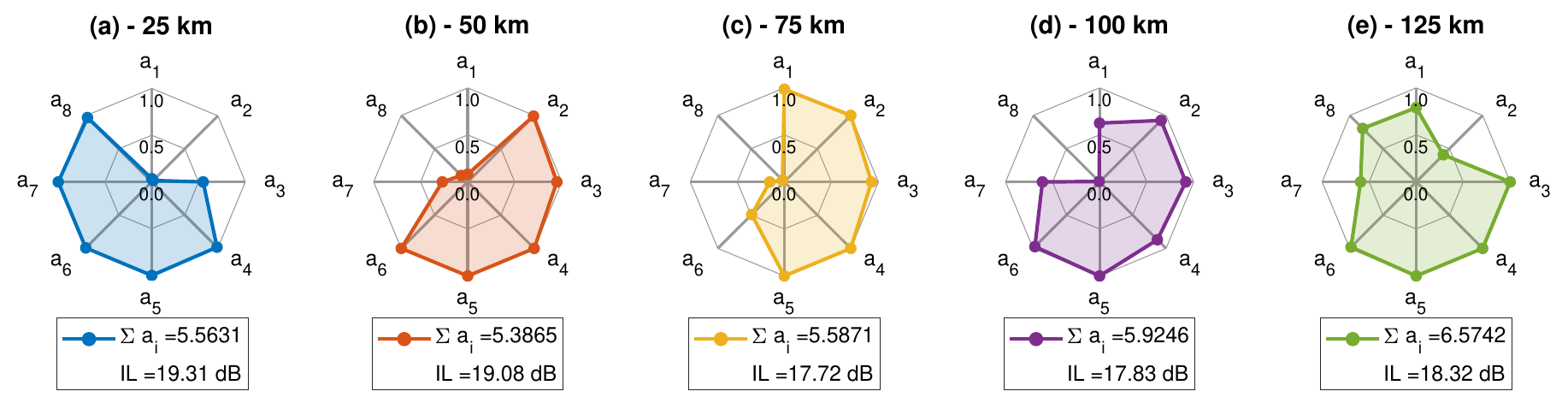}
    \caption{Optimal amplitude weights $a_i$ with $i=1,\dots,8$ obtained after the training of the PNN in \textit{Full} configuration (8 amplitude and 7 phase weights) for channel equalization at (a) 25 km, (b) 50 km, (c) 75 km, (d) 100 km, and (e) 125 km. The boxes below each diagram report the sum of the aperture levels $\Sigma_{i=1}^{N=8} a_i$ for that transmission scenario and the related insertion losses, IL.}
    \label{fig:chApertures}
\end{figure*}

\section{Results \& Discussion}
\label{sec:results}
The equalization capabilities of the PNN have been assessed with multiple training procedures in different scenarios. A first exploratory phase has been devoted to finding optimal training conditions in terms of the number of parameters and the loss function. Starting from the most general case, all 15 currents (7 PS, 8 MZI) have been used as training parameters for the PSO (\textit{full} configuration). Fig.  \ref{fig:chApertures} presents the optimal amplitude weights $a_i$, the sum of the apertures $\sum_{i=1}^{8} a_i$, and the measured insertion loss (IL) for each transmission length between 25 km and 125 km with steps of 25 km. Analogously to what has already been pointed out in \cite{staffoli2024chromaticPW}, the graphs highlight that the number of open channels ($a_i \neq 0$) increases with the fiber length. Indeed, the CD-induced temporal broadening of the transmitted symbols increases with the propagation distance, thus the PNN needs more channels to recombine pieces of optical information that are spread over a larger time window. Training results for $L=25$ km break this trend. If we consider $\sum_{i=1}^{8} a_i$ as a measure of the overall channel opening, we note that the sum results higher than the corresponding value for $L=50$ km. Indeed, for short propagation distances, the impact of cumulated CD is limited, therefore it is expected a small number of open channels for the compensation. However, limiting the number of open channels increases the IL, with a consequent degradation of the Signal-to-Noise Ratio (SNR) at the receiver. As a consequence, in the presence of reduced CD, the PNN favors an $a_i$ configuration that maintains a high SNR rather than using the minimum amount of open channels needed for the compensation. 

\begin{figure}[t]
    \centering
    \includegraphics[width=1\linewidth]{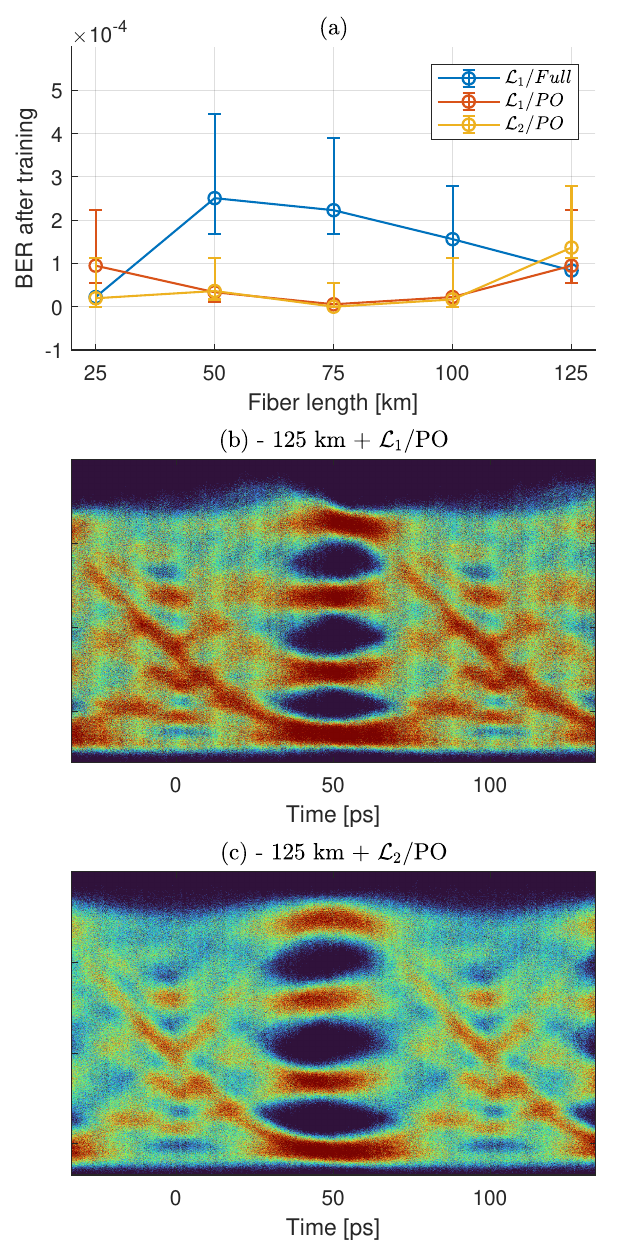}
    \caption{Comparison between different training strategies performed with PSO. (a) BER values measured after the training performed in full configuration with $\mathcal{L}_1$ (blue line), in PO configuration with $\mathcal{L}_1$ (orange line), and in PO configuration with $\mathcal{L}_2$ (yellow line) as a function of the fiber length. Each value results from the average over 20 measurements with the error bars corresponding to 68\% credible interval for the associated Poisson distribution. (b-c) Eye diagrams at the receiver after the training performed in PO configuration with (b) $\mathcal{L}_1$ and (c) $\mathcal{L}_2$ respectively.}
    \label{fig:lossfun_comparison}
\end{figure}

In a second set of measurements, the tunable parameters have been restricted to the phasors only (\textit{PO} configuration) with $a_i = 1$ for every channel. Fig.  \ref{fig:lossfun_comparison} summarizes the comparison between the different training configurations with the two tested loss functions $\mathcal{L}_1$ and $\mathcal{L}_2$. In particular, Fig.  \ref{fig:lossfun_comparison}(a) shows the BER reached after a PSO run for different fiber lengths. Each point is obtained as the average of 20 measurements with the optimal current configuration provided by the training. Each colored curve is associated with a particular combination of the loss function and the number of parameters. Since CD acts by applying a phase mask to the signal (as seen with \eqref{eq:y_inf_cperc}), we found an overall better performance achieved with PO configuration, already observed in \cite{staffoli2024chromaticPW}. The minimization of $\mathcal{L}_1/PO$ and $\mathcal{L}_2/PO$ for $L=125$ km (worst case scenario) produces the equalized eye diagrams reported respectively in Fig.  \ref{fig:lossfun_comparison}(b) and Fig.  \ref{fig:lossfun_comparison}(c). The use of $\mathcal{L}_2/PO$ allows, in general, to retrieve the symmetry of the eye, therefore it is preferred to $\mathcal{L}_1/PO$ in the rest of the paper.

\begin{figure*}[t]
    \centering
    \includegraphics[width=\linewidth]{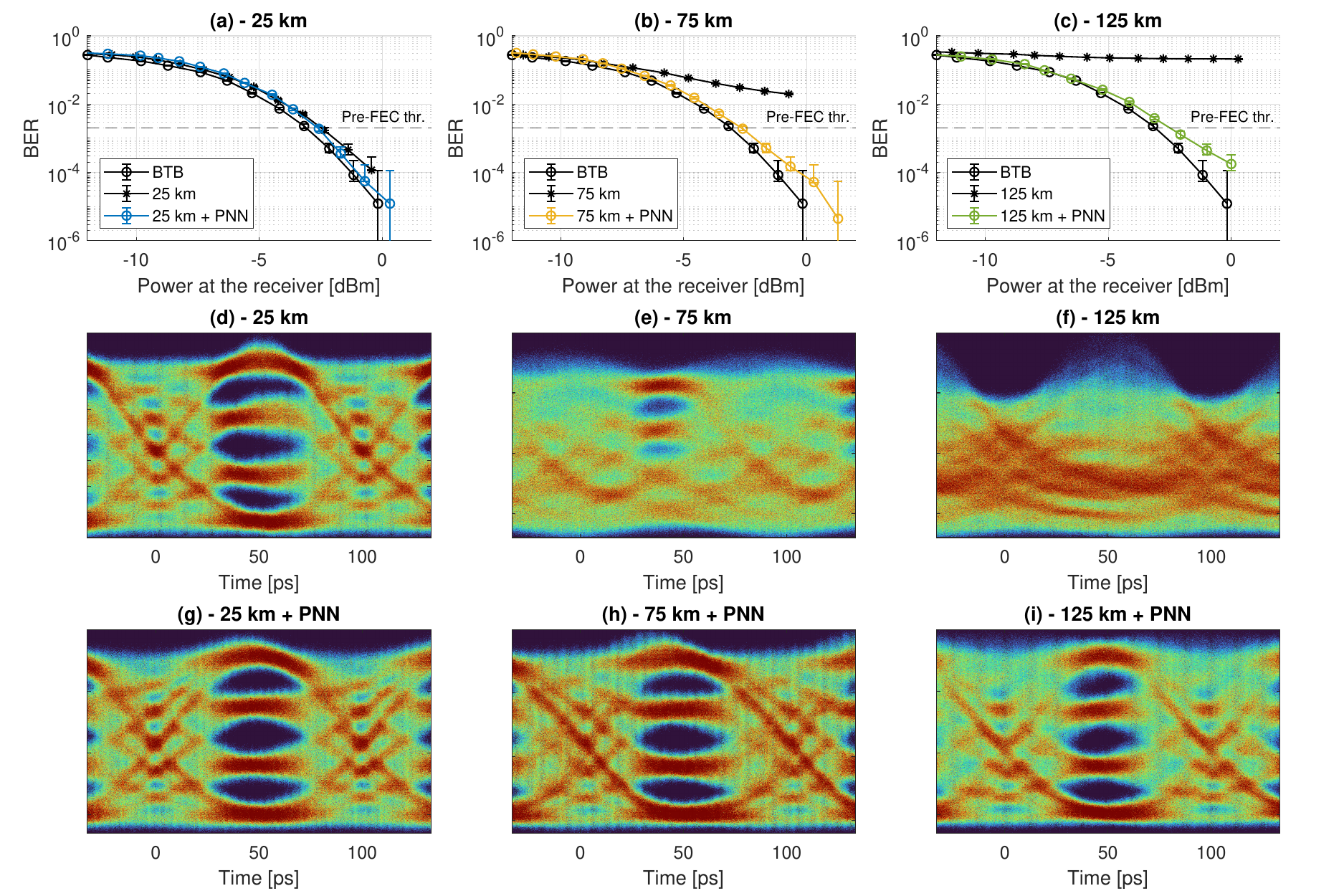}
    \caption{(a-c) BER versus Power at the receiver profiles for propagation in (a) 25 km, (b) 75 km), and (c) 125 km. The unequalized (black stars) and equalized (colored circles) are compared with the Back-to-Back performance (BTB, black circles). The training procedures have been performed with the PSO and BER values obtained as an average over 50 acquisitions and their error bars represent the 68\% credible interval for the corresponding Poisson distribution. (d-f) Unequalized eye diagrams for (d) 25 km, (e) 75 km, and (f) 125 km propagation. (g-i) Equalized eye diagrams for (g) 25 km, (h) 75 km, and (i) 125 km propagation. }
    \label{fig:bervsprx_eyes}
\end{figure*}

The equalization tests were performed for different propagation distances, from $L=25$ km to $L=125$ km, using 25 km steps. The quality of the Transmitter/Receiver performance has been assessed for each $L$ during the \textit{testing phase}. This consists of a scan over the power at the receiver (PRX), or equivalently over the Signal-to-Noise Ratio (SNR), performed by varying an optical attenuator placed in front of the end-of-line photodiode (see Fig.  \ref{fig:setup}). For each PRX, the BER was acquired as the average over 50 acquisitions. The testing phase has been performed for bare fiber transmission (unequalized case) and after each training (equalized case), finally comparing the results with the benchmark performance set by the transmission at $L=0$ km (BTB). The BER versus PRX profiles obtained for $L=25$ km, $L=75$ km, and $L=125$ km are presented in Fig.  \ref{fig:bervsprx_eyes}(a-c). The intersymbol interference generated by CD raises the unequalized BER curves with respect to the BTB condition and causes a progressive closure in the corresponding eye diagrams (Fig.  \ref{fig:bervsprx_eyes}(d-f)) as $L$ increases. The trained PNN lowers the BER curves close to the BTB performance, reaching the pre-FEC threshold of $2\times 10^{-3}$ \cite{mizuochi2006recent} in correspondence of $\text{PRX}\approx -3$ dBm, as well as the opening of the eye diagrams (Fig.  \ref{fig:bervsprx_eyes}(g-i)). 

\begin{figure}[t]
    \centering
    \includegraphics[width =1\linewidth]{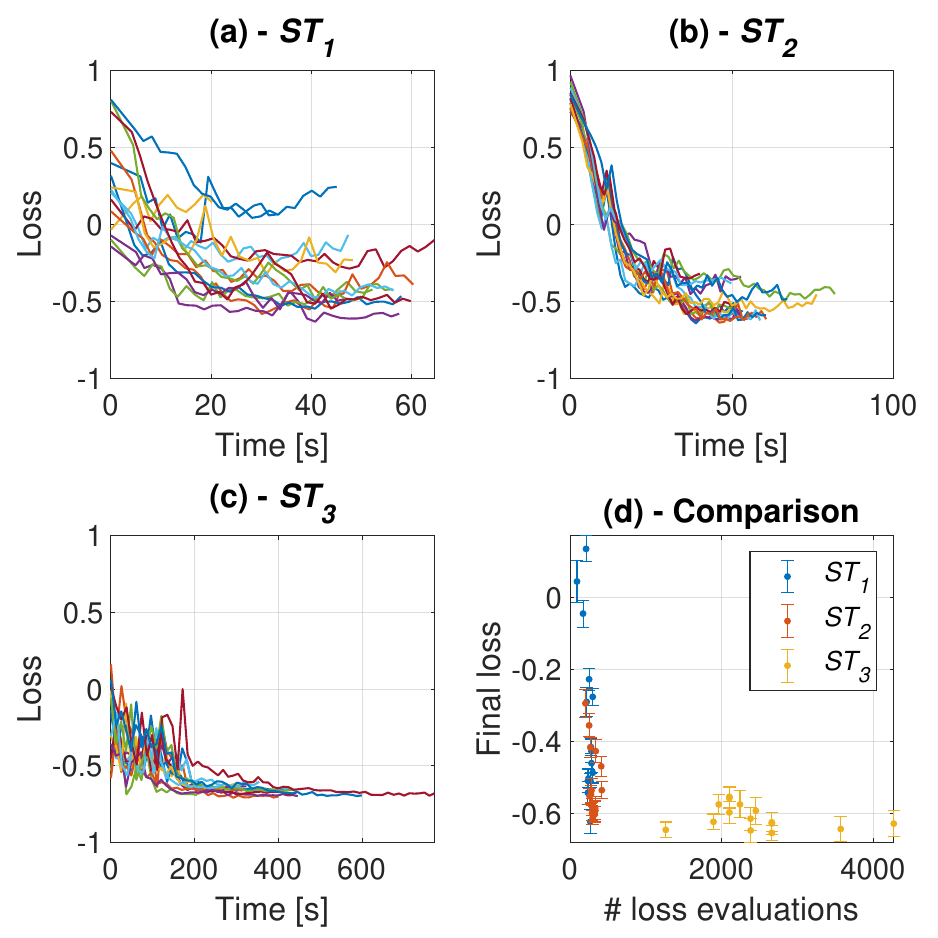}
    \caption{Comparison of different training strategies to minimize $\mathcal{L}_2$ for signal equalization after propagation in 125 km fiber. (a-c) Loss value as a function of epoch time during the training performed with strategy (a) $\mathit{ST}_1$ (Adam with random starting point), (b) strategy $\mathit{ST}_2$ (Adam with fixed starting point, and (c) strategy $\mathit{ST}_3$ (PSO). In strategy $\mathit{ST}_2$. The starting point for $\mathit{ST}_2$ corresponds to the optimal configuration (estimated with PSO) for BTB transmission. (d) Final loss function against the number of evaluations. Each value and its error bars are obtained respectively as the average and standard deviation over 20 measurements in the optimal configuration found at the end of each training. }
    \label{fig:loss_vs_niter}
\end{figure}

Then, PSO as a minimization algorithm has been compared with other strategies. Strategy $\mathit{ST}_1$ consists of a gradient descent performed by the Adam algorithm \cite{kingma2017adam} with the starting point chosen as the configuration that provides the lowest loss among 20 randomly generated ones. Similarly, the strategy $\mathit{ST}_2$ again exploits the Adam algorithm,  but the initial condition is fixed to the optimal configuration (found with PSO) that optimizes the BTB transmission. Finally, strategy $\mathit{ST}_3$ represents the standard training configuration that produced the results of Fig.  \ref{fig:bervsprx_eyes}, namely the parameters are optimized via PSO starting from a random initial condition. The optimization ends when the loss fails to improve by more than 0.02 for 10 consecutive iterations in $\mathit{ST}_1$ and $\mathit{ST}_2$ and for 15 consecutive iterations in $\mathit{ST}_3$. The final loss value after each training is evaluated as the average of 20 measures acquired with the optimal weights. Each strategy has been executed 15 times to optimize the equalization of 125 km fiber with the intent to assess its performance in terms of speed, precision, and repeatability. Fig. \ref{fig:loss_vs_niter}(a-c) reports the best loss at every epoch as a function of time for the three strategies, while Fig.  \ref{fig:loss_vs_niter}(d) compares the final loss values against the number of loss function evaluations spent in the minimization process. Strategy $\mathit{ST}_1$ fails to reach an optimal solution in most cases, while strategy $\mathit{ST}_2$ is more consistent in reaching low loss values. The PSO-based strategy $\mathit{ST}_3$ achieves the lowest loss values on average, but the number of iterations needed for the minimization is increased by one order of magnitude compared to $\mathit{ST}_1$ and $\mathit{ST}_2$. The Adam algorithm is thus faster than the PSO, but the reliability strongly depends on a proper choice of the starting point. Our experiments prove that a favorable starting configuration for an equalization problem is the one that minimizes the loss function at BTB, namely the one that makes the PNN as transparent as possible to the signal. PSO guarantees convergence to a near-to-optimal configuration at the cost of increased complexity.

\begin{figure}
    \centering
    \includegraphics[width =1\linewidth]{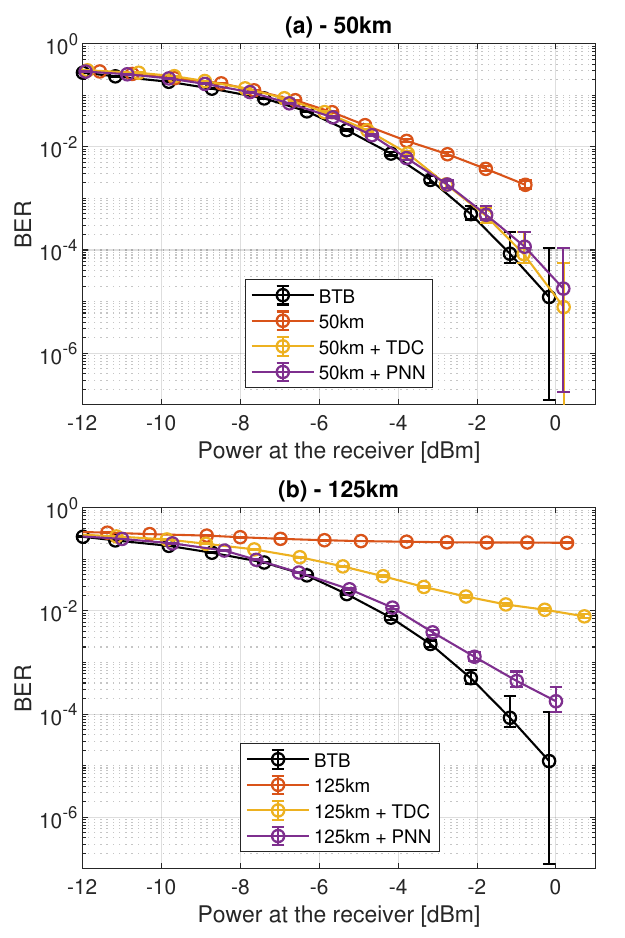}
    \caption{BER versus Power at the receiver profiles for equalized and unequalized transmission in (a) 50 km and (b) 125 km. The profiles refer to transmission in BTB (black), bare fiber (orange), and compensated with TDC (yellow) and PNN (purple). }
    \label{fig:tdc_comparison}
\end{figure}

The performance of the PNN has been compared against a commercial Tunable Dispersion Compensator (TDC) based on fiber-Bragg gratings\footnote{TeraXion ClearSpectrum$^\text{TM}$ Model: TDCMB-C000-J.}. The device features an insertion loss of \SI{5}{\dB} and a dispersion range of $\pm\SI{900}{\pico\s/\nano\m}$, corresponding to a complete equalization of CD accumulated in a 50 km propagation with a 10 Gbaud signal. In the experimental setup described in Fig.  \ref{fig:setup}, the PNN was replaced by the TDC. Since the PNN and the TDC have different insertion losses, it is necessary to adjust the input power to the new device to restore the same fiber launch power condition. In Fig.  \ref{fig:tdc_comparison} the BER curves obtained at 50 km and 125 km using the TDC or the PNN are compared. The two devices are equally effective in retrieving the signal after 50 km of fiber, while the PNN demonstrates a larger dispersion range with good performance up to 125 km.

\begin{figure}
    \centering
    \includegraphics[width =1\linewidth]{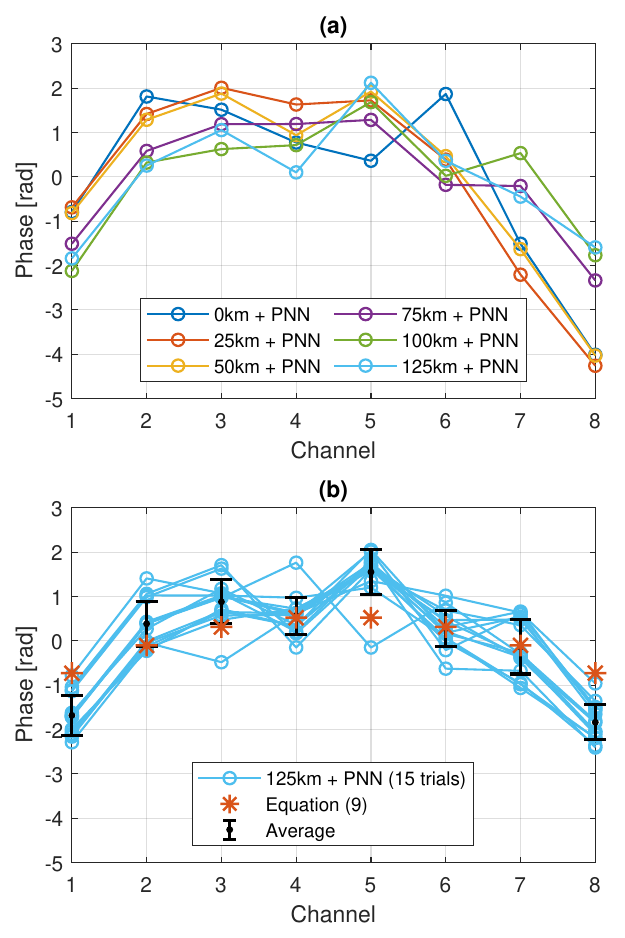}
    \caption{(a) Optimal phases resulting from PSO for different transmission distances. (b) Optimal phases resulting from 15 different training for $L=125$ km using PSO compared with the optimal phases for a perception with infinite channels and an infinitesimal delay unit (see \eqref{eq:phases_ideal}).}
    \label{fig:phases}
\end{figure}

As mentioned, a set of current values is the result of the training. This set can be translated into a set of phase shifts according to
\begin{equation} \label{eq:alpha_beta}
    \phi_i = \alpha_i I_i^2 + \beta_i,
\end{equation}
with $i=1,\dots,8$. These values appear in \eqref{eq:y_cperc} and determine the optical response of the PNN. The coefficients $\alpha_i$ and $\beta_i$ in \eqref{eq:alpha_beta} are experimentally characterized for every microheater. The phases $\phi_i$ for each channel are measured relative to the channel with no delay since the corresponding heater is left unconnected (see Appendix \ref{sec:experimental}). Fig. \ref{fig:phases}(a) shows the phases of the PNN in $PO$ configuration trained to compensate for different lengths of optical fiber. Fig.  \ref{fig:phases}(b) shows the phases of the PNN, again in $PO$ configuration, obtained for 15 different minimization procedures to equalize the propagation in 125 km. The phases measured in a particular trial are adjusted by $\pm 2\pi$, and the average value is subtracted to superimpose the curves obtained in different trials. This is feasible because the response of the PNN is invariant under a constant phase shift over all channels. \textcolor{black}{ Fig. \ref{fig:phases}(b) shows that the phases resulting from a repeated minimization show a trend that is not fully compatible with \eqref{eq:phases_ideal}, especially in the first and last channels}. The variability in the training outcomes is a consequence of the PSO algorithm, which involves randomness in the initialization and evolution of the swarm, as well as statistical fluctuations in the loss evaluation. In addition to this, the possibility of ending up in local minima cannot be ruled out. Let us also stress that the configuration described by \eqref{eq:phases_ideal} was derived under the assumption of a PNN with a large number of channels and small $\Delta t$. The PNN under test does not satisfy these assumptions. Moreover, \eqref{eq:phases_ideal} neglects the insertion loss of the PNN. A larger insertion loss implies a  smaller SNR at the receiver, which in turn translates into a narrower separation of the optical levels and a larger value of $\mathcal{L}_2$ (see Fig.  \ref{fig:hist_example}). Therefore, the configuration that minimizes the loss consists of a trade-off between the compensation of CD and the reduction of the insertion loss. \textcolor{black}{In the PNN under test, the application of the ideal phases for $L=125$ km rather than those obtained after the optimization results in 2.5 orders of magnitude increase in the BER. A more detailed analysis is reported in Appendix \ref{sec:performance}.}

\begin{figure}[t]
    \centering
    \includegraphics[width =1\linewidth]{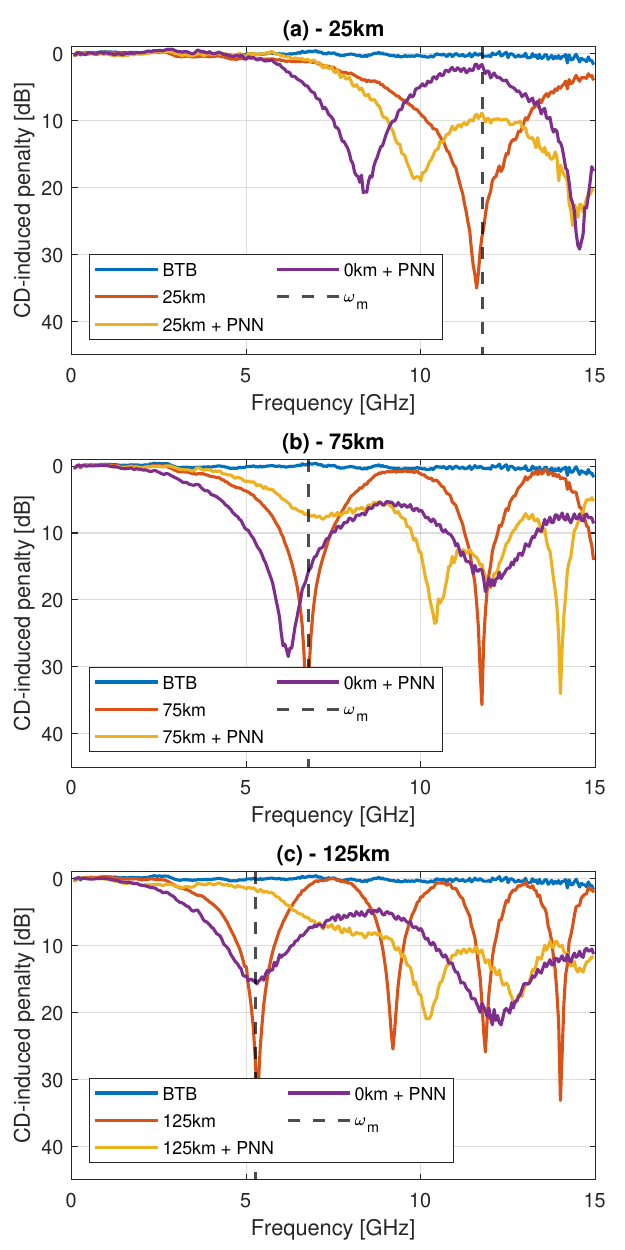}
    \caption{CD-induced penalty of the optical link for (a) 25 km, (b) 75 km, and (c) 125 km propagation. The BTB profile (blue) is reported for a comparison with the unequalized (orange), and equalized (yellow) profiles. The frequency response of the trained PNN is reported (violet) along with the $\omega_m$ parameter (dashed vertical line) for each configuration.}
    \label{fig:freq_resp}
\end{figure}

Fig. \ref{fig:freq_resp} shows the CD-induced penalty of the equalized (yellow) and unequalized (red) 25 km, 75 km, and 125 km long optical link. The CD-induced penalty at BTB (blue) is reported for comparison. In violet is the CD-induced penalty measured after removing the fiber span from the link and keeping just the PNN. \textcolor{black}{ Note that the yellow curves do not result from the summation of the red and purple ones. Indeed, they describe the frequency response of the whole system formed by the PNN and the fiber span.}. In the 25 km case, bare fiber propagation does not significantly alter the power penalty within the bandwidth of the modulation (10 GHz). In this case, the PNN is useless, and actually, its action moves the first notch of the power penalty around \SI{10}{\giga\Hz}. On the other hand, at 75 km and 125 km, the differences between BTB and bare fiber profiles are significant within the bandwidth of the transmitted signal. In these cases, the PNN reduces the CD-induced penalty by shifting the first notch back to higher frequency values. One can observe that when the fiber span is removed from the link, keeping just the trained PNN, the first minimum of the CD-induced penalty is approximately where it would be with the fiber in place and without the PNN. This observation suggests that the frequency response of the trained PNN approximates the inverse of the optical fiber frequency response within the bandwidth of the signal \textcolor{black}{ so that their product is equal to 1, as expected following \eqref{eq:Hideal}.}

\begin{figure*}
    \centering
    \includegraphics[width =1\linewidth]{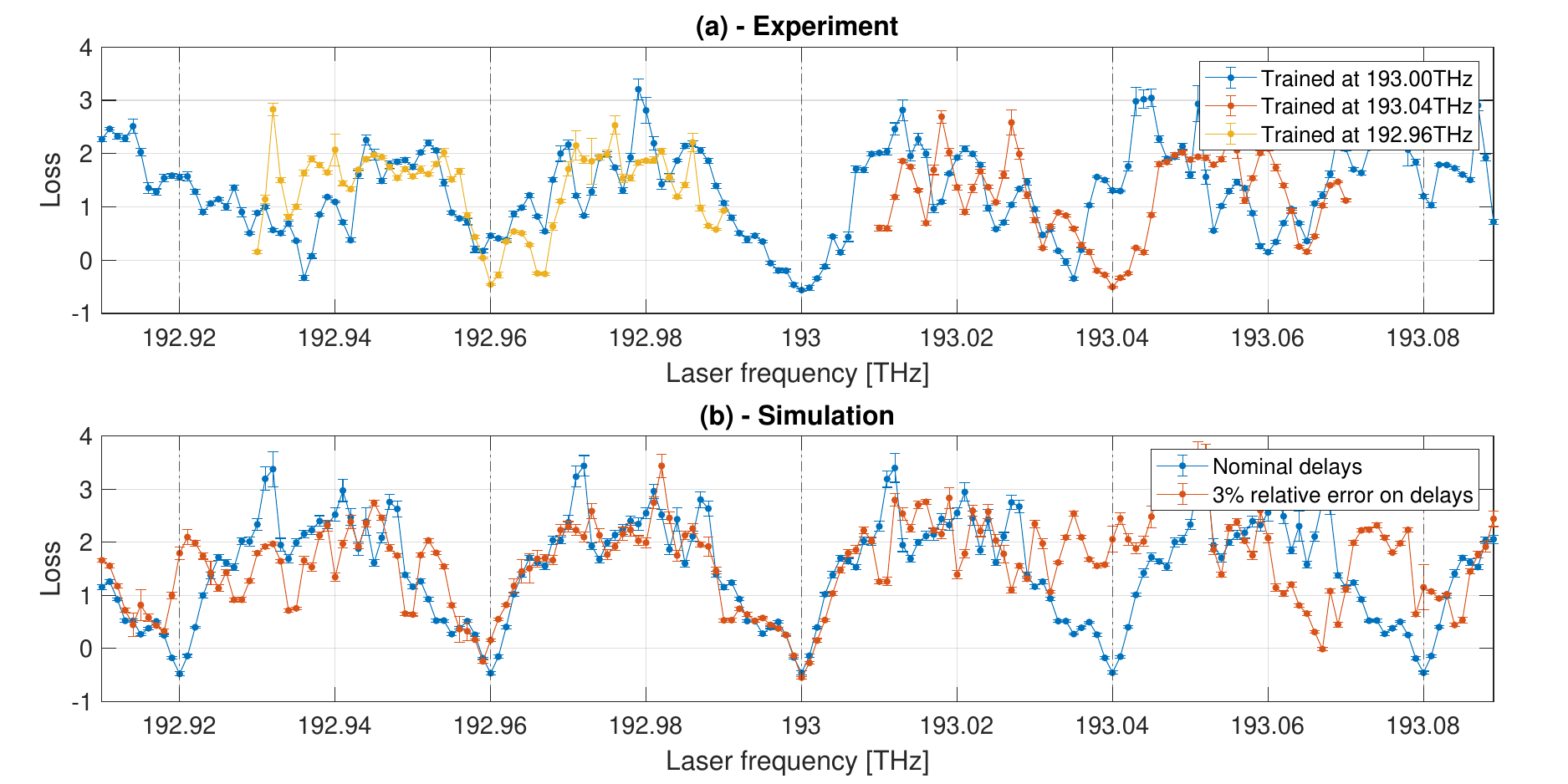}
    \caption{Loss as a function of the optical carrier frequency. (a) Experiment with the PNN trained at \SI{193}{\tera\Hz} (blue), \SI{193.04}{\tera\Hz} (orange) and \SI{192.96}{\tera\Hz} (yellow). Lower loss values indicate better equalization. Each point and its error bar are obtained respectively as the average and standard deviation of 20 repeated measures. As the laser frequency is shifted from the one used during the training, the PNN fails to achieve low loss values. (b) Simulation assuming the nominal spiral lengths (blue) or a 3\% relative error on those values. The results confirm that manufacturing imperfections can spoil the periodicity of the response.}
    \label{fig:loss_vs_freq}
\end{figure*}

To test the applicability of our PNN to a DWDM link, we tested its equalization capabilities by varying the laser frequency at the transmitter. The experiment is performed with 125 km of optical fiber using the trained configuration of the PNN at $f_0=193$ THz. The weights of the PNN are left unaltered throughout the frequency sweep, which covers the interval $f_o \pm 90$ GHz with steps of 1 GHz. The end-of-line optical filter follows the peak frequency during the sweep. The blue line in Fig. \ref{fig:loss_vs_freq}(a) shows how the loss function $\mathcal{L}_2$ changes with the transmitted frequency. Each point and its error bar in the curve are obtained respectively as the average and the standard deviation of 20 experiments. It can be seen that as the signal frequency changes by a few GHz from the training frequency $f_0$ the loss increases, namely the PNN becomes ineffective at compensating CD. This is explained by the fact that the phases accumulated by the signal along the spirals change with the optical frequency. Since the phase shifts induced by the micro-heaters remain the same, the optimal interference condition for the equalization is lost. The same observation is made by training the PNN at \SI{193.04}{\tera\Hz} (orange line) and \SI{192.96}{\tera\Hz} (yellow line). The Free Spectral Range (FSR) $\Delta \lambda$ of the PNN is given by
\begin{equation}
    \Delta \lambda = \frac{c}{n_g L_C} \approx \SI{40}{\giga\Hz},
\end{equation}
where $n_g$ is the group index of the PNN Silicon waveguides. Therefore, we expected that the optimal weights found by training the PNN at \SI{193}{\tera\Hz} would also work at \SI{192.96}{\tera\Hz} and \SI{193.04}{\tera\Hz}. However, this is not the case. \textcolor{black}{ A plausible reason lies in fabrication imperfections. Indeed, the periodicity in frequency is lost if the delays are not multiples of the same unit as a result of variations in the spiral lengths or, more plausibly, in the group velocity.  The electron beam lithography technique used to trim the waveguide induces local changes in the track width (Line Width Roughness), which are perceived as variations in the average group index at different frequencies. The role played by uncertainty in the delay is confirmed by performing a simulation of the experiment (see Fig. \ref{fig:loss_vs_freq}(b)).} The periodicity is well visible in the simulation with nominal values (blue line) and disappears when errors in delays are considered.

\section{Comparison with other approaches}
\label{sec:comparison}
\renewcommand\tabularxcolumn[1]{>{\Centering}m{#1}}
\begin{table*}[t]
    \centering
    \begin{tabular}{c|
    >{\centering \arraybackslash}m{3.2cm}
    >{\centering \arraybackslash}m{1.5 cm}
    >{\centering \arraybackslash}m{1cm}
    >{\centering \arraybackslash}m{2.5 cm}
    >{\centering \arraybackslash}m{1.8cm}
    >{\centering \arraybackslash}m{2 cm}
    >{\centering \arraybackslash}m{2 cm}
    }
    \hline
    \textbf{Ref.}   &   \textbf{Layout} &	\textbf{Platform}	&	\textbf{Bitrate}	&    \textbf{Encoding}	&	\textbf{Fiber}	&	\textbf{Dispersion range}	&	\textbf{WDM}   \\
        &   &	&	\textbf{(Gbps)}	&   &   \textbf{(km)}	&	\textbf{(ps/nm)}	&	\textbf{(GHz)}		\\
    \hline	
    \cite{brodnik2018extended}	&	MZI-based 10 stages O-FIR&	SiN & 53	&	PAM4	&	40	&	$\pm$500 	&	100 (exp.)		\\
    \cite{sorianello2018todc}	& MR-based 3 stages O-FIR&	Si&	100	&	PolMux NRZ	&	30	&	$-$540	&	100 (sim.)		\\
    \cite{liu2022todc}	&	MR-based 7 stages O-FIR&		Si& 25	&	NRZ	&	40	&	$-$720	&	100 (sim.)		\\
    \cite{staffoli2023equalization}	&	4-tap O-FIR	&	Si&	10	&	NRZ	&	125	&	$\pm$2200	&	80 (theo.)		\\
    \cite{nguyen2010tunable}	&	16-tap O-FIR	&	PLC&	40	&	NRZ	&	5.5	&	$\pm$100	&	100 (theo.)		\\
    \cite{marciano2024alloptical}	&	8-tap O-FIR	&	Si&	20	&	OFDM	&	75	&	$\pm$1320	&	40 (theo.)		\\
    \cite{liu2023adaptive}	&	4-tap O-FIR + DSP	&	SiN/LNOI &	28	&	QPSK-4	&	30	&	$-$570	&	28 (theo.)		\\
    \cite{argyris2018photonic}	&	O-RC	& VCSEL+OF&	25	&	NRZ	&	45	&	$-$810+NL	&	-		\\
    \cite{sackesyn2021fiber}	&	O-RC	&	Si&	32	&	NRZ	&	25	&	$-$450+NL	&	-		\\
    \cite{sozos2021photonic}	&	O-RC + 51-tap FIR	&	Si&	224	&	PAM4	&	60	&	$-$1080	&	-		\\	\hline										
      this work	&	8-tap O-FIR	&	Si&	20	&	PAM4	&	125	&	$\pm$2200	&	40 (theo.)		\\\hline
    \end{tabular}
    \vspace{2pt}
\caption{Comparison between different techniques for CD compensation. O-FIR: Optical Finite Impulse Response filter, MR: Microring, RC: Reservoir Computing, NL: Nonlinearities, SiN: Silicon Nitride, Si: Silicon, PLC: Planar Lightwave Circuit, LNOI: Lithium Niobate On Insulator, VCSEL: Vertical Cavity Surface Emitting Laser, OF: Optical Fiber, WDM: Wavelength Division Multiplexing. WDM entrance refers to the spacing between channels that the device in a static configuration manages to compensate.}
\label{tab:comparison}
\end{table*}

Table \ref{tab:comparison} proposes a comparison between several implementations of photonic devices used for CD compensation. Over the years, several integrated photonic filters have been proposed for this scope. In \cite{brodnik2018extended} an integrated filter consisting of 10 cascaded unbalanced MZIs was proved to equalize a 53.125Gb/s PAM4 signal over 40 km of optical fiber. The device can be tuned in the range $\pm500$ ps/nm (considering a 15 GHz bandwidth) by adjusting a single voltage and can equalize up to 4 WDM channels with a spacing of 100 GHz. A similar performance was achieved by filters consisting of cascaded ring resonators \cite{sorianello2018todc, liu2022todc}. In general, the maximum dispersion that cascade filters can compensate grows with the number of stages that, however, leads to a larger insertion loss.

Optical FIR filters with complex coefficients have been studied for CD compensation as well. These devices can be classified as transversal-form or lattice-form filters \cite{takiguchi2006optical}. In a transversal-form filter, the signal goes through a $1\times N$ splitter, then a phase shift is applied to each branch and finally a $N\times1$ combiner produces the output. Our device falls into this category. A lattice-form filter consists of a cascade of MZIs and delay lines. This architecture allows for lower transmission losses but setting the desired configuration is not straightforward \cite{jinguji1995synthesis}. Transversal-form filters differ in the way the splitter/combiner is implemented. It is possible to use a tree of y-branches/MMI (as in our previous work \cite{staffoli2023equalization}) followed by one MZI per tap (as in this work) or a tree of MZI as in \cite{nguyen2010tunable,liu2023adaptive}. It is more convenient to implement the splitter as a tree of MZIs so that the input power can be distributed over the branches with different weights without losses. On the other hand, if the splitter is a tree of y-branches, power is lost depending on the aperture of the MZIs.

Different approaches have been followed in the application of photonic neural networks to signal recovery in optical communication. One that has gained considerable attention is Reservoir Computing (RC), in which the signal is fed into a non-trainable recurrent network with randomly connected non-linear nodes called reservoir \cite{vanDerSande2017advances}. The only trainable part of the net is a linear regressor (or classifier) that receives as inputs some of the nodes of the reservoir. In \cite{argyris2018photonic}, the recovery of the transmitted sequence from the distorted signal is framed as a classification task and it is demonstrated that a reservoir comprising a semiconductor laser with a delayed feedback loop can successfully tackle it. Masking is used to increase the dimension of the reservoir, implying the necessity of sampling more points per symbol, which is impractical at high baud rates. In \cite{sackesyn2021fiber} equalization is achieved with an integrated passive photonic reservoir (with a swirl topology and 16 nodes) in which the non-linearity is provided by the photo-detectors that probe the nodes. This means that the linear stage must be implemented as part of a DSP, even though the reservoir allows for a simpler and more power-efficient DSP that performs only linear operations. In \cite{sozos2021photonic} the reservoir consists of two parallel optical filters with feedback. The idea behind this design is that a higher computational power can be achieved by sampling different spectral slices of the signal. 

Compared to other solutions, our device proposes a competitive approach to the problem of CD equalization. Its compact design offers a wide compensation range with great adaptability to different modulation formats \cite{kabir2008ofdm}. The layout is easily scalable compared to other technologies, as discussed in the next section. It allows maintaining the signal processing entirely at the optical level (except for the training phase), with minimized latency and power consumption compared to other approaches. Manufacturing imperfections affect its applicability to DWDM and increase the insertion losses.

\section{Conclusion}
\label{sec:conclusion}
\begin{figure*}[t]
    \centering
    \includegraphics[width =1\linewidth]{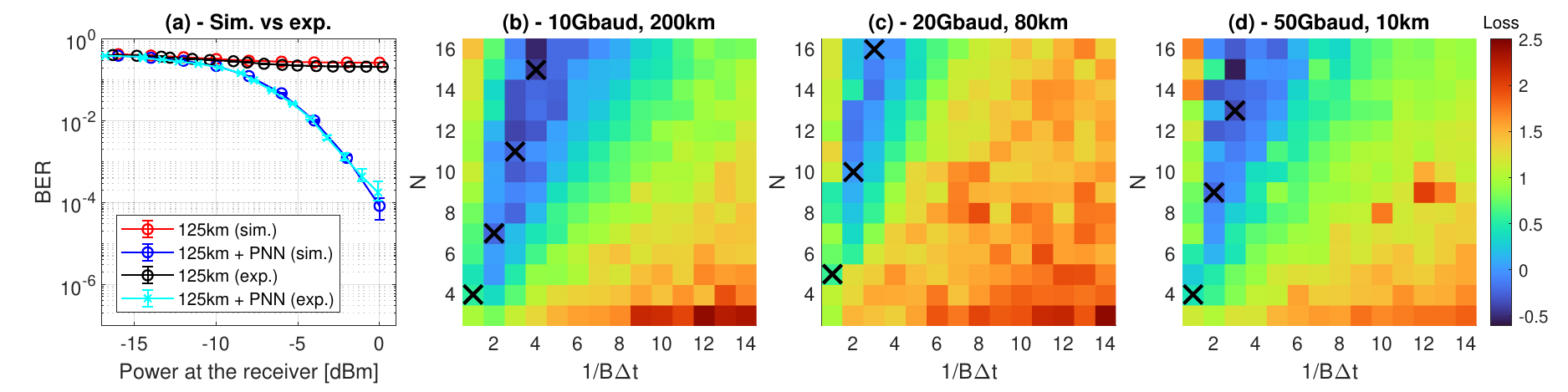}
    \caption{(a) Comparison between the experimental BER curves and the ones obtained from a simulation for a PAM4 10Gbaud signal and a transmission distance of 125km. (b-c) Variation of the simulated final loss value for different combinations of the number of channels $N$ and the delay unit $\Delta t$ for a given baudrate $B$. On the x-axis, the product $(1/B) \times \Delta t$ indicates the fraction of baud duration occupied by the delay unit. The crosses indicate the optimal combinations of $N$ and $\Delta t$ predicted by Equation \eqref{eq:ntaps}.}
    \label{fig:loss_heatmap}
\end{figure*}

We demonstrated that an 8-channel delayed complex perceptron effectively equalizes a 10 Gbaud PAM4 signal transmitted through up to 125 km of optical fiber (18.5 dB average insertion loss). The action of the device can be described as the discrete convolution between the input signal and a finite sequence of tunable complex weights. Two methods for finding the optimal configuration of these weights, PSO and Adam, were compared: PSO is more precise and accurate than Adam with the drawback of an increased computational cost. The proposed PNN proved to be as effective as a commercial TDC in compensating CD while requiring less power consumption during the operation ($\sim 250$ mW against few W). The device's effectiveness is confirmed by the fact that it flattens the CD-induced penalty within the bandwidth of the transmitted signal. Due to manufacturing imperfections, the 40 GHz periodicity in the response of the PNN predicted by the ideal model was not observed, limiting its applicability to WDM schemes.

We simulate the effect of structural parameters, namely the number of channels $N$ and the delay unit $\Delta t$, on the PNN performance. As shown in Fig. \ref{fig:loss_heatmap}(a), the equalized and unequalized BER curves at 125 km obtained from the simulation are compatible with the experimental ones, proving the code's reliability.
Different training procedures of the PNN for various combinations of $N$ and $\Delta t$ are performed via PSO, each by evaluating the loss in the trained configuration. For consistency, the loss is always evaluated with the same SNR at the receiver, and the bandwidth of the transmission system is scaled proportionally to the bandwidth baud rate. Three scenarios are considered: a 10 Gbaud PAM4 signal propagating through 200 km of fiber (Fig. \ref{fig:loss_heatmap}(b)), a 20 Gbaud PAM4 signal propagating through 80 km of fiber (Fig. \ref{fig:loss_heatmap}(c)) and a 50 Gbaud PAM4 signal propagating through 10 km of fiber (Fig. \ref{fig:loss_heatmap}(d)). Every scenario has a set of parameters for which the PNN reaches low loss values, indicating an effective signal equalization. The results show that the best performance is obtained in proximity to the parameters that satisfy \eqref{eq:ntaps}, marked by crosses in Fig. \ref{fig:loss_heatmap}. Equation \eqref{eq:ntaps} expresses the fact that the product $N\Delta t$, namely the time window observed by the PNN, must be comparable with $\beta_2L\Delta\omega$, the broadening caused by CD. In general, it is observed that a more significant number of channels, thereby a smaller delay unit, ensures better performance. Indeed, a PNN with more degrees of freedom can better approximate the inverse transfer function of the optical fiber. \textcolor{black}{ The simulations demonstrate that a time-delay complex perceptron with less than 16 channels is capable of equalizing a 50 Gbaud signal up to 10 km of transmission distance.  However, similarly to other integrated photonic devices, the scalability comes at the cost of a larger footprint, higher power consumption by the microheaters, enhanced thermal cross-talk, and a longer training duration. Many efforts are being put into conciliating increased complexity with performance \cite{shekhar2024roadmapping}. Focusing on our case of interest, future-generation devices could feature an on-chip integrated optical amplifier (e.g., an SOA) for loss reduction and enhanced versatility (studies on the feasibility in terms of power consumption are ongoing). In parallel, the problem of thermal cross-talk and power inefficiency generated by microheaters in MZIs can be mitigated by implementing electro-absorption modulators (EAM). Both these improvements are available in the next-generation InP devices at our disposal, which will be tested soon.}

\textcolor{black}{We foresee that future-generation devices will also represent a valid alternative to coherent transmission in access and metro applications. On single-span links, the use of the PNN device will reduce the number of components required in the link, thus simplifying the network and reducing the costs. On multi-span links, using the PNN device increases the quality of the transmission by fully compensating chromatic dispersion or removing the residuals. Migrating the equalization to the optical domain mitigates typical DSP-related problems such as power consumption, latency, and costs. Compared to coherent systems, the innovative PNN approach still features some drawbacks, such as high insertion losses, poor performances in WDM networks, and low receiver sensitivity typical of IMDD. Future generation devices will implement new features (on-chip SOA and EAM), making the device more competitive with respect to algorithm-based alternatives and more prone to scalability. We plan to perform tests for nonlinear effects equalization and coherent modulation formats to demonstrate its versatility and advantages.}


%

\appendices
\section{Experimental setup}
\label{sec:experimental}
\begin{figure*}[t]
    \centering
    \includegraphics[width =1\linewidth]{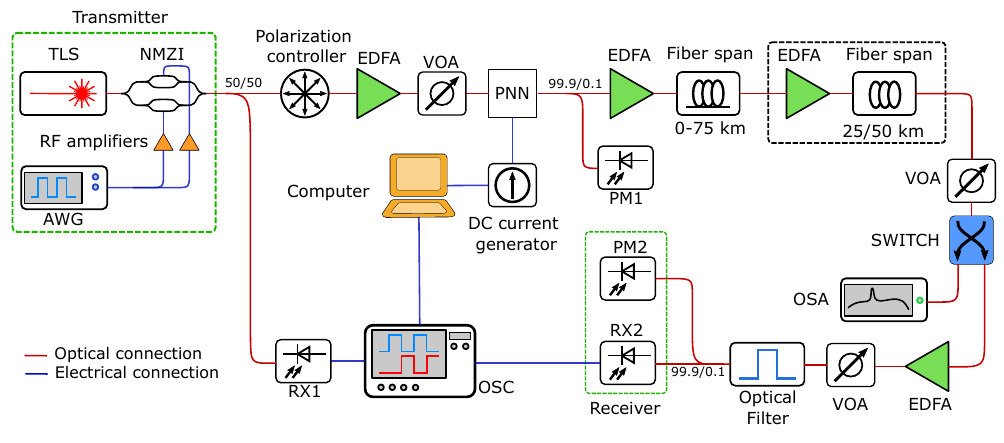}
    \caption{Experimental setup. TLS: Tunable Laser Source, NMZI: Nested Mach-Zehnder Interferometer, RF: Radio Frequency, AWG: Arbitrary Waveform Generator, EDFA: Erbium-Doped Fiber Amplifier, VOA: Variable Optical Attenuator, PNN: Photonic Neural Network, RX: Receiver, OSA: Optical Spectrum Analyzer, PM: Power Meter, OSC: Oscilloscope.}
    \label{fig:setup}
\end{figure*}

Fig. \ref{fig:setup} depicts a schematic of the complete experimental setup. A CW tunable laser source (TLS) operating at 1550 nm is modulated as a 4-level Pulse Amplitude Modulated (PAM4) signal. This is software-generated and based on a periodic Pseudo-Random Binary Sequence (PRBS) of order 10 and period $2^{10}$. The pattern is generated by a 30 GHz-bandwidth Arbitrary Waveform Generator (AWG) connected to two 20 GHz-bandwidth RF amplifiers, which drive a Nested Mach-Zehnder Interferometer (NMZI). A 50/50 Fiber Optic Coupler sends half of the optical power to a 20 GHz-bandwidth fast photodiode (RX1), where the input signal is acquired for reference. The other half of the optical power proceeds to a polarization controller, which is needed since the PNN device is polarization-sensitive (see Section \ref{sec:dcp_layout}). \textcolor{black}{An Erbium-Doped Fiber Amplifier (EDFA) and a Variable Optical Attenuator (VOA) regulate the input power to the PNN device so that it never exceeds 3 dBm in order not to damage the device and not to trigger nonlinear effects in Silicon. The PNN is placed before the fiber to eliminate polarization-dependent losses.} The coupling to the chip occurs through butt coupling via tapered fibers held by 3-axis stages. The 0.1\% of the collected optical power is to a Power Monitor (PM1) for alignment monitoring. The PNN processes the optical signal accordingly to \eqref{eq:y_cperc}, with the weights driven by a DC current generator with 16 independent channels. Since the PNN optical response is defined for one of the phase weights $e^{j\phi_i}$, the PS associated with the null-delay channel (Channel 1) is left unconnected. Therefore, only 15 of the 16 channels of the DC current generators are used to tune the PNN device, while the remaining channel is used as a triggering signal for the oscilloscope (measurement procedure described below). The PNN device is soldered to a Printed Circuit Board (PCB) and electrically connected via wire bonding. A Copper plate in direct thermal contact with the PNN device is placed underneath the PCB and connected to a Temperature Controller for thermal stabilization. 

The remaining 99.9 \% of the PNN optical power transmission is sent to an EDFA to compensate for the PNN insertion losses ($\sim 20$ dB with all the currents switched off). Then, the optical signal is coupled to a first fiber span with variable length between 0 and 75 km with steps of 25 km. Larger propagation distances are reached by implementing another 25 km or 50 km fiber span preceded by another EDFA to prevent Optical Signal-To-Noise Ratio (OSNR) degradation. Standard SM G.652D fibers have been used (0.2 dB/km losses), with the input optical power to each span never exceeding 2 dBm to prevent the triggering of nonlinear effects. The fiber link transmitted output optical power is attenuated by a VOA and, via an optical switch, is addressed alternatively to an Optical Spectrum Analyzer (OSA) or to a final amplification stage which compensates for fiber losses. Subsequently, another VOA regulates the amount of optical power sent to a 30 GHz optical bandwidth filter and then to the receiver. A fiber splitter routes part of the signal towards a second Power Meter (PM2, 0.1\%) and part to another 20 GHz-bandwidth fast photodiode (RX2, 99.9\%). Both RX1 and RX2 are connected to an 80 GSa/s oscilloscope (OSC) with a 16 GHz bandwidth. 

The OSNR level at 0 dBm at RX2 is maintained at $(44 \pm 1)$ dB (0.1-nm resolution bandwidth) for the measurement process. Since the optical gain of the end-of-line EDFA is left unaltered for the whole experiment duration, the mentioned condition at the receiver reflects into $(34 \pm 1)$ dB OSNR (0.1-nm resolution bandwidth) and $(-14 \pm 1)$ dBm optical power at 1550 nm measured at the OSA. This is chosen as the reference working point for the measurements performed in each transmission scenario. When the PNN device is inserted, this condition is set while keeping all currents at 0 mA in the device. Some deviations from this working point continuously occur during the training, since the insertion loss of the device significantly changes with the current configuration set in each specific measure (these changes cannot be monitored directly by the OSA due to the too-low duty cycle of its measure). Similarly, the trained PNN may induce deviations from the reference working point. However, a noise characterization of the experimental setup revealed that these variations in PRX and OSNR minimally impact the SNR profile at the receiver. The consistency between the measurements performed with and without the PNN is thus preserved.

Table \ref{tab:ch_losses} reports channel losses $k_i$ entering in \eqref{eq:y_cperc} obtained after the characterization of the PNN device. 

\begin{table}[h!]
    \centering
    \begin{tabular}{cc|cc}
        \hline
        Channel \#  &    Loss $k$ [dB]  &    Channel \#  &    Loss $k$ [dB]     \\
        \hline
        1   &   -19.0   &   5   &  -21.4      \\
        2   &   -15.5   &   6   &  -16.0      \\
        3   &   -14.8   &   7   &  -18.0      \\
        4   &   -14.7   &   8   &  -20.0      \\
        \hline
    \end{tabular}
    \vspace{2pt}
    \caption{Results of the PNN characterization procedure in terms of channel losses $k_i$ entering in \eqref{eq:y_cperc}.}
    \label{tab:ch_losses}
\end{table}

\section{Performance comparison between ideal and non-ideal implementations}
\label{sec:performance}
\begin{figure*}[t]
    \centering
    \includegraphics[width=1\linewidth]{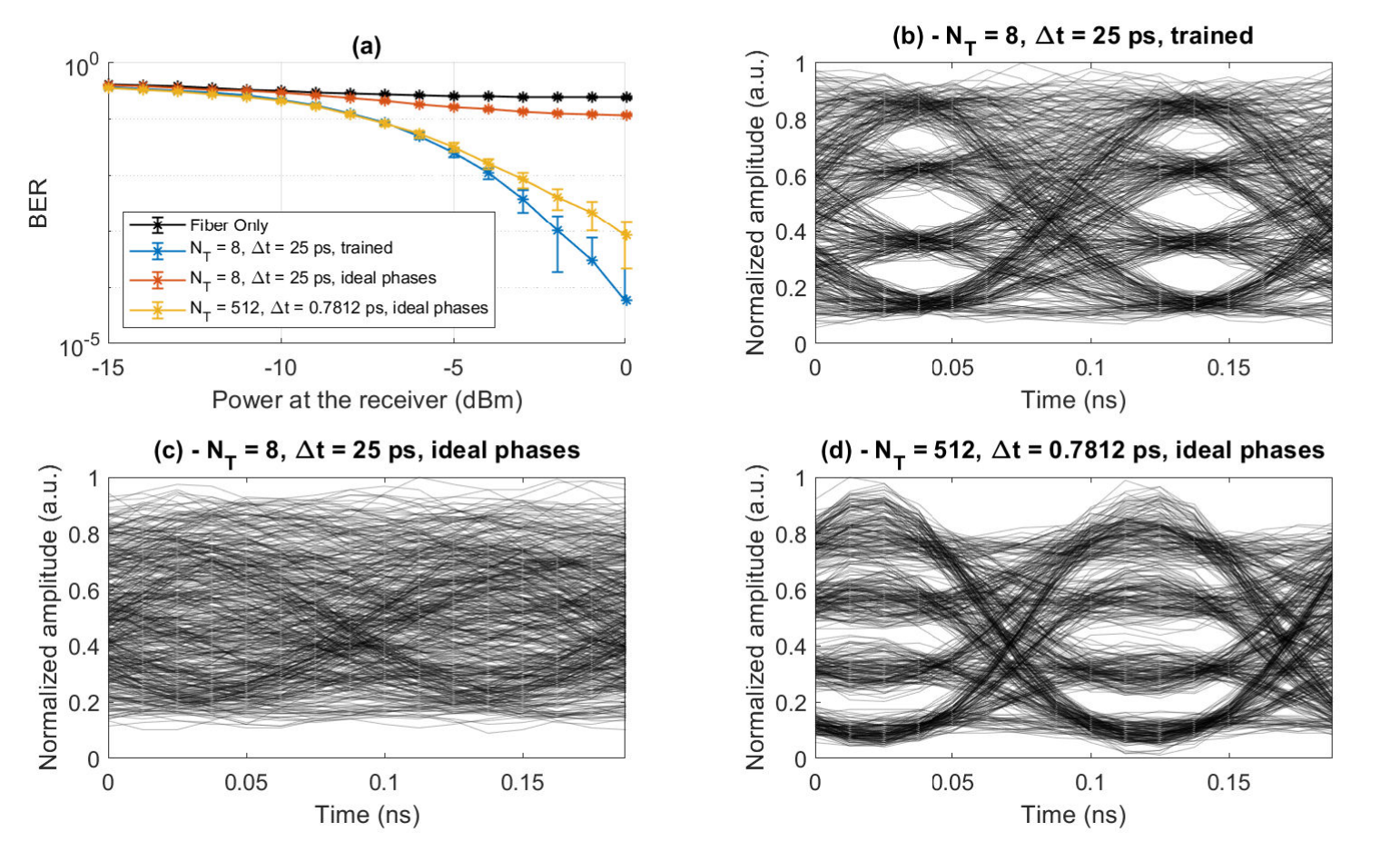}
    \caption{Simulation results for different PNN configurations in terms of $N_T$, $\Delta t$, and applied phase weights. (a) BER versus Power at the receiver profiles and (b-d) eye diagrams at the receiver for different PNN layouts and applied phases specified (a) in the legend and (b) in the titles, respectively.}
    \label{fig:ideal_vs_nonideal_ber_eye}
\end{figure*}

\textcolor{black}{
Following the discussion about Fig. \ref{fig:phases}, here we provide a detailed discussion about the differences between the non-ideal PNN device (finite $N_T$ and finite $\Delta t$) and the ideal behavior predicted by \eqref{eq:phases_ideal}. The comparison is carried out on a simulative level using the same code exploited for the results of Fig. \ref{fig:loss_heatmap}. The benchmark performance is set with a simulated PNN with $N_T = 8$ taps and a delay unit $\Delta t = 25$ ps. The PNN is trained via Particle Swarm Optimizer (PSO) in Phase Only (PO) configuration to equalize CD after 125 km propagation, saving the eye diagram at the receiver and BER versus power at the receiver (PRX) profile. The results have been compared with those produced respectively by the configurations $N_T = 8$, $\Delta t = 25$ ps, and $N_T = 512$, $\Delta t = 0.7812$ ps applying the ideal phases derived from \eqref{eq:phases_ideal}. In particular, the last configuration adheres more to the ideal specifications under which \eqref{eq:phases_ideal} was derived. The comparison of the BER versus PRX profiles for the three cases and the corresponding eye diagrams at 0 dBm at the receiver is reported in Fig. \ref{fig:ideal_vs_nonideal_ber_eye}. As expected, the lowest BER profile is obtained with the trained PNN, namely when the phase weights are optimally adapted to the selected transmission scenario, resulting in the opened eye diagram of panel (b). The same PNN layout applied with the ideal phases does not perform similarly, bringing the BER profile close to the unequalized case and resulting in the degraded eye diagram of panel (c). Applying the ideal phases brings an advantage as soon as the PNN layout adheres to the ideal case ($N_T \rightarrow \infty$, $\Delta t \rightarrow 0$). In this case, an improvement in the eye diagram aperture is observed in panel (d). }

\section*{Acknowledgment}
European Research Council (ERC) under the European Union’s Horizon 2020 research and innovation programme (grant agreement No 788793, BACKUP and No 963463, ALPI). 
The authors would like to thank Mattia Mancinelli and Davide Bazzanella for the chip design and Stefano Biasi for the fruitful discussions.

\ifCLASSOPTIONcaptionsoff
  \newpage
\fi



%
\bibliographystyle{IEEEtran}
\bibliography{bibliography}

%







\end{document}